\documentstyle[12pt]{article}
\newcommand{\PSbox}[3]{\mbox{\rule{0in}{#3}\includegraphics{#1}\hspace{#2}}}
\begin{document}
\newcommand{\pl}[1]{Phys.\ Lett.\ {\bf #1}\ }
\newcommand{\npb}[1]{Nucl.\ Phys.\ {\bf B#1}\ }
\newcommand{\prd}[1]{Phys.\ Rev.\ {\bf D#1}\ }
\newcommand{\prl}[1]{Phys.\ Rev.\ Lett.\ {\bf #1}\ }

\addtolength{\textheight}{0.4 in}

\title{Signatures of Technicolor Models \\ 
with the GIM Mechanism$^*$}

\author{Witold Skiba \vspace{0.4cm} \\ 
{\it Center for Theoretical Physics} \\
{\it Laboratory for Nuclear Science and Department of Physics} \\
{\it Massachusetts Institute of Technology} \\ 
{\it Cambridge, MA 02139, USA} }
\footnotetext[1]{This work is supported in
part by funds provided by the U.S. Department of Energy under
cooperative research agreement DE-FC02-94ER40818.\hfill}
\date{}

\maketitle

\vspace{-3.5in}
\rightline{ \begin{tabular}{l}
MIT-CTP-2505  \\ 
January 1996  \\ 
\end{tabular} }
\vspace{ 3.7in}

\begin{abstract}
We investigate the production and the decays of pseudo-Goldstone bosons (PGBs)
predicted by technicolor theories with the GIM mechanism (TC-GIM)\@. The
TC-GIM models contain exotic fermion families that do not interact under weak
$SU(2)$, but they do have color and hypercharge interactions. These fermions 
form PGBs, which are the lightest exotic particles in the TC-GIM models.
The spectrum of PGBs consists of color octets, leptoquarks and neutral
particles. The masses of leptoquarks and color octets depend on a free
parameter---the scale of confining interactions. Characteristic for TC-GIM
models is a very light ($\approx \!\!1$~GeV) neutral  particle with anomalous
couplings to gauge boson pairs. We show how current experiments constrain
the free parameters of the models. The best tests are provided by the 
$pp\rightarrow TT$ and $e^+e^- \rightarrow P^0 \gamma$ reactions. Experiments
at LHC and NLC can find PGBs of TC-GIM models in a wide range of parameter
space. However, TC-GIM models can be distinguished from other TC models only
if several PGBs are discovered. 
\end{abstract}

\thispagestyle{empty}
\newpage
\setcounter{page}{1}

\section{Introduction}
Models of weak interactions based on the technicolor idea~\cite{technicolor}
need a mechanism which communicates the symmetry breaking to the quarks
and leptons. One way of coupling the technifermions to the quarks and leptons 
is the introduction of extended-technicolor (ETC) 
interactions~\cite{techniETC}, which generate fermion masses. However, 
simple extended-technicolor models suffer unacceptable flavor-changing 
neutral currents. More limitations on viable technicolor theories are 
imposed by measurements of the electroweak parameters\cite{techniSTU}.

An interesting solution to these problems are models which incorporate
the Glashow-Iliopoulous-Maiani (GIM) mechanism \cite{GIMmechanism}. The 
first technicolor models that used GIM mechanism to avoid unacceptably large
flavor-changing neutral currents were the composite technicolor standard
models \cite{CTSM}. These models realize the GIM mechanism by separating
the ETC interactions into several ETC groups. There are separate ETC 
groups for the left-handed fermion fields, the right-handed up quarks,
and the right-handed down quarks. Such construction introduces a large global
symmetry associated with quark flavor. This flavor symmetry is the
essence of the GIM mechanism. Breaking of the global symmetry is responsible
for the fermion masses and the quark mixing---the existence of the
Kobayashi-Maskawa matrix. However, the composite technicolor 
models presented in Ref.~\cite{CTSM} were toy models of weak 
interactions, since the models did not incorporate leptons. 

Realistic technicolor models with the GIM mechanism  were described
in Refs.~\cite{themodel} and \cite{physfromvacuum}. We will refer to these
models as technicolor-GIM models (TC-GIM)\@. Not only do the TC-GIM models
avoid trouble with the flavor-changing neutral currents, but they also
limit the number of technifermion doublets to one, thereby avoiding conflict 
with precise electroweak measurements~\cite{techniSTU}. A noticeable feature
of these models is the presence of exotic light fermions. From the point of
view of a model builder, the most difficult task is to create a model
with an appropriate pattern of breaking flavor and gauge symmetries.
The symmetry breaking is achieved by introducing numerous heavy fermion
fields and gauge bosons. The light fermions that were mentioned before 
exist in the TC-GIM models only to cancel certain anomalies. In QCD-like
models the light fermions seem to be a necessary ingredient. Since the light
fermions are a necessary feature of TC-GIM models, their signatures are
the best place to test and study this kind of models.

In this paper we explore the phenomenological consequences of the light 
fermion sector. The light fermions transform under the ETC groups and also 
some additional confining interactions. The scale of confining interactions, 
depending on a particular model, can be from tens to hundreds of GeV\@. Below 
the confinement scale, there are pseudo-Goldstone bosons (PGBs) in the 
particle spectrum whose constituents are the light fermions.  The PGBs
are the lightest exotic particles in the spectrum of the TC-GIM models.
While present experiments put a lower bound on the scale of the confining 
interactions, future experiments may find signatures of the PGBs.
The light fermions, constituents of the PGBs, do not transform under the
ordinary $SU(2)_L$ gauge group. Their interactions with the quarks and
leptons are mediated by the ETC gauge bosons. 

PGBs in TC-GIM have a different origin than PGBs in other types of TC models.
Usually, PGBs are associated with chiral symmetry breaking by the technicolor
group. In TC-GIM models, PGBs arise from dynamical breaking of symmetry by
some new gauge interactions. Of course, the TC group also breaks chiral 
symmetries of technifermions. However, in TC-GIM, there is only one doublet
of technifermions and the Goldstone bosons become longitudinal degrees of
freedom of the $W^\pm$ and the $Z$\@. TC-GIM models, like any other TC models,
have techni-$\rho$ mesons, but these are heavier than the technicolor scale
and more difficult to observe than PGBs. Therefore, the PGBs are the best 
test of TC-GIM models. If PGBs are discovered and some of their properties 
are known, it will be possible to distinguish between different TC scenarios.

In the next section, we begin with a short introduction to the TC-GIM models 
and explain various possible realizations of the light fermion sectors. In 
Section~\ref{sec:spectrum}, we present the spectrum of the PGBs and their
couplings to ordinary particles. Section~\ref{sec:rates} contains
the discussion of the PGB phenomenology. We leave some remarks about the 
scales of the interactions that confine the light fermions until 
Section~\ref{sec:scales}. 

\section{TC-GIM models}
\label{sec:models}

\begin{figure}[!ht]
\PSbox{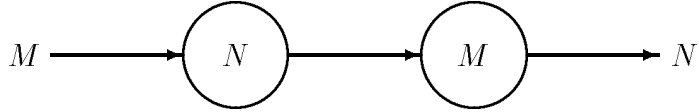 hscale=100 vscale=100 hoffset=-70 
       voffset=-620}{13.7cm}{2cm} 
  \caption{An example of moose diagram. The circles represent $SU(N)$ and
    $SU(M)$ gauge groups, the lines represent fermion fields. Line endings
    without circles represent global symmetry groups --- SU(M) and SU(N).}
\label{fig:moose}
\end{figure}

The basic building blocks of the TC-GIM models are $SU(N)$ gauge groups and 
chiral fermions. Because many gauge groups are necessary, a special notation
``moose notation'' is helpful in describing these models~\cite{toolkit}.
An $SU(N)$ gauge group is represented by a circle. Fermions in the 
(anti-)fundamental representation are depicted by an (in-)out-going line. 
A fermion line connecting two circles represents fermions transforming under 
both gauge groups depicted by the circles. A line whose one end is not
connected to any circle indicates that the fermions transform under a 
gauge group and a global symmetry group. A graphic illustration of these 
ideas is presented in Fig.~\ref{fig:moose}, where a simple moose diagram 
represents three fermion fields transforming under two global groups and 
two gauge groups. When referring to fermion fields, we will label the 
fermions $[NM]$, where $N$ and $M$ stand for $SU(N)$ and $SU(M)$ groups 
under which the fermions transform.

All gauge groups are guaranteed to be anomaly-free by having the same
number of in-going and outgoing fermion lines. This way, all fermions
transforming under a given gauge group form a vector representation,
which is anomaly-free. The requirement of anomaly cancellation is
how the light fermions find their way into TC-GIM models. The ETC gauge 
groups would not be anomaly-free without these additional fermions.

\begin{figure}[!ht]
  \PSbox{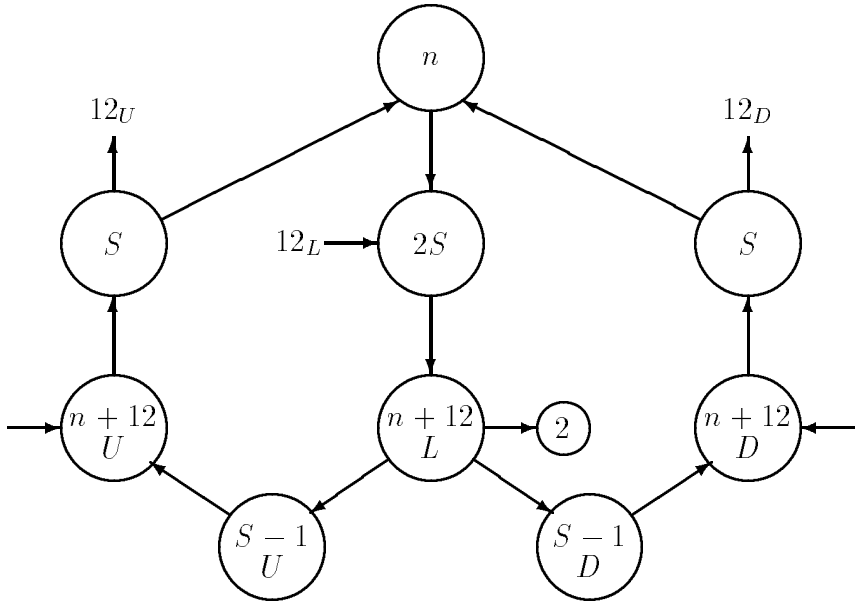 hscale=100 vscale=100 hoffset=-70 
         voffset=-420}{13.7cm}{8.5cm} 
  \caption{The full TC-GIM model with a low-scale PGB sector.}
  \label{fig:model}
\end{figure}

The structure of a TC-GIM model described in Ref.~\cite{themodel} is
illustrated in Fig.~\ref{fig:model}. The gauge groups are ordered 
by their scale, with the lowest scale groups at the bottom of the
moose. There are two $SU(S-1)$ groups labeled U and D\@. The fermions
transforming under these groups are the light fermions forming PGBs.
The single $SU(2)$ group is the familiar group of weak interactions.
The model contains three separate ETC groups: one for the left-handed 
quarks and leptons, one for the right-handed down quarks and charged
leptons, and one for the right-handed up quarks and neutrinos. These are
the three $SU(N+12)$ groups. $N$ is the number of techni-colors,
while $12$ is the number of left-handed doublets of quarks and leptons.
The fact that there are right-handed neutrinos in the model does not present
any problem. There exist several plausible mechanisms to ensure small masses
of the neutrinos~\cite{themodel}. The two $SU(S)$'s together with the $SU(2 S)$
and $SU(N)$ groups at the top of the moose are the highest scale groups.
They break the ETC groups and merge several $SU(N)$ subgroups into one
technicolor $SU(N)$. As usual, the technifermion condensates break 
the weak $SU(2)$.  We will not elaborate on the details of the TC-GIM model
building, instead we refer the reader to Refs.~\cite{themodel} and 
\cite{physfromvacuum}. The $[n+12_L,2]$ fermions include all left-handed
quarks and leptons, and also one technifermion doublet. The right-handed
quarks and leptons are contained in the $[n+12_U,1]$ and $[n+12_D,1]$
fermion lines.

We will focus on the PGBs formed from fermions transforming under 
$SU(S-1)$ groups. The $SU(S-1)$ groups have to become strongly interacting 
at some scale, otherwise there would be massless or very light fermions 
(with masses comparable to those of leptons and quarks) present in the 
particle spectrum. The scale of the $SU(S-1)$ interactions is not
related to the technicolor scale. In most technicolor models, the lightest
exotic particles are PGBs formed by technifermions. Such PGBs form when
the TC group dynamically breaks chiral symmetries of the technifermions.
The same process gives masses to the electroweak gauge bosons, thus the 
scale of these PGBs is related to the scale of electroweak symmetry breaking.
The situation is different in TC-GIM models. PGBs are created when the 
$SU(S-1)$ interactions form fermion condensates. Therefore, the scale at which
PGBs form in TC-GIM models is a free parameter but there are both upper
and lower bounds on its value. It cannot be too low in order to avoid conflict
with experiments. On the other hand, too large a scale would give too large 
contributions to lepton and quark masses, or be so high as to 
upset the hierarchy of symmetry breaking. We postpone the discussion 
of these problems to Sec.~\ref{sec:scales}, after we discuss the 
phenomenological constraints. 

There are several possibilities for constructing the light-fermion sectors.
The one illustrated in Fig.~\ref{fig:model} is characterized by a relatively
low scale of the $SU(S-1)$ interactions. We will later show that this 
specific model cannot accommodate heavy PGBs. In particular, the lightest
leptoquarks have to be lighter than approximately 50~GeV, therefore
the low scale model is ruled out. From now on, we will suppress the structure 
of the models irrelevant to PGB study, showing only parts of the moose we 
are interested in: the ETC groups, the ordinary fermions and the light-fermion 
sector. The up and down sectors of the moose are identical so it is enough 
to describe any one of the two sectors.

\begin{figure}[!t]
  \PSbox{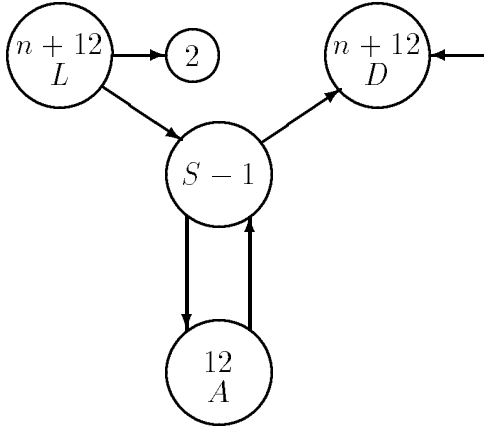 hscale=100 vscale=100 hoffset=-60 
         voffset=-495}{13.7cm}{5.5cm} 
  \caption{High-scale model of the PGB sector.}
  \label{fig:highscale}
\end{figure}

A model similar to the one presented in Fig.~\ref{fig:model} was also 
presented in Ref.~\cite{themodel}. We illustrate the relevant part in 
Fig.~\ref{fig:highscale}. This model is characterized by a higher scale
of the $SU(S-1)$ interactions. The difference between this model
and the one introduced before is the addition of $[12_A,S-1]$ and 
$[S-1,12_A]$ fermion fields. When the $SU(S-1)$ interactions become strong,
we assume that the $[12_A,S-1]$ fermions form condensates with the
$[S-1,n+12_D]$ fermions. Likewise $[n+12_L,S-1]$ fermions form condensates 
with the $[S-1,12_A]$. This is clearly not the only possibility. 
If the $[n+12_L,S-1]$ fermions formed a condensate with the $[S-1,n+12_D]$
fermions, this model would not be very different from the model depicted
in Fig.~\ref{fig:model}.

\begin{figure}[!hb]
  \PSbox{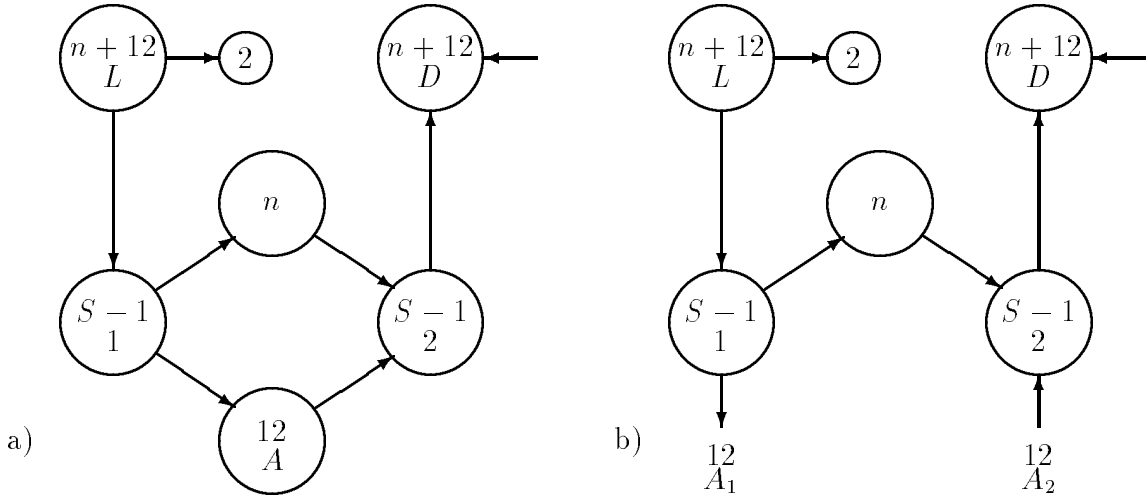 hscale=100 vscale=100 hoffset=-70 
         voffset=-470}{13.7cm}{6.6cm} 
  \caption{High-scale models of PGB sectors without vacuum-alignment 
           ambiguity.}
  \label{fig:highscaleab}
\end{figure}

It is possible to construct models whose light fermion sectors have
unambiguous vacuum state. Such a model was introduced in 
Ref.~\cite{physfromvacuum}, whose relevant part we reproduce in
 Fig.~\ref{fig:highscaleab}a. A variation of such a model is 
illustrated in Fig.~\ref{fig:highscaleab}b. There are two separate 
$SU(S-1)$ groups in these models. For simplicity, we assume that both 
groups are characterized by the same scale, but it would not present 
any difficulties to deal with different scales. In these models, the 
$[12_A,S-1_2]$ fermions form condensates with the $[S-1_2,n+12_D]$ fermions,
and similarly $[n+12_L,S-1_1]$ with $[S-1_1,12_A]$. We will refer to the 
model illustrated in Fig.~\ref{fig:model} as the low-scale model, because the
scale of $SU(S-1)$ interactions in this model has a strong upper bound. All 
other models presented here avoid that upper limit, so we will refer to them
as the high-scale models.

In the limit where there is no lepton mixing and the K-M matrix is diagonal,
there are separate conserved lepton and quark numbers for each flavor
of quarks and leptons. The exotic fermions transform under the global
$U(1)$ symmetries associated with the flavor numbers. Therefore, we
can attribute flavor to the exotic fermions and, consequently, refer
to them as `exotic electron' or `exotic top quark', etc. The exotic
and usual fermions carry the same quantum numbers of color and electric 
charge, however the exotic fermions are always singlets of the 
weak $SU(2)_L$.   
 
\section{Spectrum and couplings}
\label{sec:spectrum}
We will now describe the PGBs formed below the scale of the $SU(S-1)$
interactions. We first enumerate the PGBs and estimate their masses.
We then derive the couplings of the PGBs to ordinary quarks, leptons
and gauge bosons. In what follows, we use chiral Lagrangian techniques.
Such a description is valid for the momenta of the PGBs smaller than the
chiral symmetry breaking scale of the $SU(S-1)$ interactions.

In the low-scale model, the PGBs are associated with the breaking of the 
$SU(12)\times SU(12)$ flavor symmetry, where the $SU(12)$ groups are 
subgroups of $SU(n+12)$ ETC groups. The $SU(12)^2$ global symmetry breaks
down to $SU(12)$, so there will be $12^2-1=143$ PGBs. The symmetry group 
of the high-scale models is $SU(12)^4$. This group is dynamically broken 
to $SU(12)^2$, doubling the number of the PGBs. Not all bosons remain in 
the particle spectrum, some may be eaten by the Higgs mechanism. The 
$SU(12)^2$ or $SU(12)^4$ are not exact symmetries. Once we take into 
account symmetry breaking interactions, the PGBs will acquire masses.

We describe here the PGBs formed in the sector associated with the down 
quarks. Precisely the same analysis applies to the up sector.
The $SU(12)$ symmetries contain the ordinary color $SU(3)$ gauge group.
The exotic quarks and leptons have the same charge assignment as 
the quarks and leptons. Therefore, there exist two flavor $SU(3)$ symmetries
embedded in the $SU(12)$. These symmetries are associated with independent
rotations in the $(d,s,b)$ and $(e,\mu,\tau)$ spaces, where we use
the same letters to describe the ordinary or exotic fermions. 

Let $Q_c$ and $L$ denote flavor $SU(3)$ triplets:
\begin{displaymath}
  Q_c=\left( \begin{array}{c} d_c \\ s_c \\ b_c \end{array} \right),\
  L=\left( \begin{array}{c} e \\ \mu \\ \tau \end{array} \right)
\end{displaymath}    
and $\lambda^i$ denote the Gell-Mann matrices, normalized such that
$Tr(\lambda^i \lambda^j) = \frac{1}{2} \delta^{ij}$. The upper
index refers to flavor space; the lower one to color space.

We classify the PGBs according to the embedding of the $SU(3)_{color}
\times SU(3)_Q \times SU(3)_L $ symmetry in the $SU(12)$ group. The
spectrum of PGBs is a straightforward generalization of the spectrum
in the old one-family technicolor model~\cite{FarhiSusskind}.
The 143 PGBs can be written as the following combinations of fields
\begin{equation}
 \label{eq:PGBs}
 \begin{array}{ll}
  \theta^i_a \sim \sqrt{2} \bar{Q} \gamma_5 \lambda_a \lambda^i Q, & \\
  \theta_a \sim \frac{1}{\sqrt{3}} \bar{Q} \gamma_5 \lambda_a Q, & \\
  T_c^i \sim \sqrt{2} \bar{Q_c} \gamma_5 \lambda^i L, &
  \bar{T_c^i} \sim \sqrt{2} \bar{L} \gamma_5 \lambda^i Q_c, \\
  T_c \sim \frac{1}{\sqrt{3}} \bar{Q_c} \gamma_5 L, &
  \bar{T_c} \sim \frac{1}{\sqrt{3}} \bar{L} \gamma_5 Q_c, \\
  \Pi^i \sim \frac{1}{2} \left( \bar{Q} \gamma_5 \lambda^i Q + 
  \bar{L} \gamma_5 \lambda^i L \right), & \\
  P^i \sim \frac{1}{2\sqrt{3}} \left( \bar{Q} \gamma_5 \lambda^i Q - 3 
  \bar{L} \gamma_5 \lambda^i L \right), & \\
  P^0 \sim \frac{1}{6\sqrt{2}} 
  \left( \bar{Q} \gamma_5 Q - 3 \bar{L} \gamma_5 L \right), &
 \end{array}
\end{equation}
where $i,a=1,\ldots,8$ and $c=1,2,3$. Whenever a flavor or color matrix
is omitted, it should be understood that the identity matrix is present 
in the relevant space.

The 64 $\theta^i_a$ bosons and 8 $\theta_a$ are color octets. The 48
$T_c^i, \bar{T_c^i}$ and the 6 $T_c,\bar{T_c}$ are color triplets. 
The $T$'s carry both quark and lepton numbers. We will refer to them as
leptoquarks. There are also 17 color singlet states, which are the 
$\Pi^i$, $P^i$ and $P^0$. Color singlet and octet states do not carry 
electric charges, the leptoquarks have charge $-\frac{2}{3} e$ 
($+\frac{2}{3} e$ for the leptoquarks in the up sector).
The spectrum of the high-scale models is a simple replication of the
spectrum just described.

In order to estimate masses of the PGBs, we need to itemize terms that 
break the $SU(12)$ global symmetries. In the low-scale model there is
an explicit mass term for the exotic fermions. The mass term originates
from multi-fermion operators~\cite{themodel}. Let
\begin{equation}
  \label{eq:nonlinear}
  \Sigma = \exp{(2 i T_a \pi^a / f_{S-1})}
\end{equation}
be the nonlinear representation of the PGBs where $T_a$ are the $SU(12)$ 
matrices. Let $L$ and $D$ be the $SU(12)$ subgroups of the ETC groups 
$SU(N+12)_L$ and $SU(N+12)_D$ under which ordinary fermions transform 
linearly
\begin{equation}
  \psi_L \rightarrow L \psi_L \; {\rm and} \; d_R \rightarrow D d_R. 
\end{equation}
Then $\Sigma$ transforms in the following manner
\begin{equation}
  \Sigma \rightarrow L^\dagger \Sigma D.
\end{equation}
The mass matrix for the exotic fermions is related to the mass matrix
for the quarks and leptons: $M = m \left(\frac{f_{ETC}}{v_{TC}}\right)^2$,
where $f_{ETC}$ is the scale of the ETC interactions, $v_{TC}$ is the value
of the technicolor condensate, which is approximately 250 GeV while $m$ is 
the mass matrix for the down quarks and charged leptons.
The lowest-order contribution to the PGBs masses comes from the 
term
\begin{equation}
\label{eq:mass}
  v_{S-1}^3 tr(\Sigma^\dagger M) + h.c.,
\end{equation}
which gives the following mass-squared matrix:
\begin{equation}
 (\Delta m_{ab})^2 = 8 \frac{v_{S-1}^3}{f_{S-1}^2} tr\left( T_a T_b M \right).
\end{equation}

Color and electromagnetic interactions also break the $SU(12) \times SU(12)$
symmetry. The standard approach in computing the color and electromagnetic
contributions is to rescale the electromagnetic mass splitting in the
$\pi^{\pm} - \pi^0$ system~\cite{emmass}. For the charged pions the leading 
effect comes from a one-photon exchange. An exchange of one gluon is similar
in structure, except for different coupling constant and some $SU(3)$ group
factors. Therefore, the contribution to the color octet (triplet) masses
can be related to the pion mass difference
\begin{equation}
  \frac{(\Delta m_\theta)^2}{m_{\pi^{\pm}}^2 - m_{\pi^0}^2} = 
  \left(\frac{f_{S-1}}{f_\pi}\right)^2 
  \frac{\alpha_{QCD}(f_{S-1})}{\alpha_{em}} \,3 \; \ \ (\frac{4}{3}),
\label{eq:scaleuppi}
\end{equation}
where the factor of 3 applies to the octet pseudos and $\frac{4}{3}$ 
to the triplets. The numerical value of this contribution to the masses
of the color octet PGBs is
\begin{equation}
  \Delta m_\theta \approx \left( \frac{f_{S-1}}{15 {\rm GeV}} \right) 
  45\, {\rm GeV}.
\label{eq:octetmass}
\end{equation}
Electromagnetic contributions to the leptoquark masses can be computed
in the same manner. Of course, instead of group theory factors there
is a factor of $\frac{4}{9}$, which is the charge squared. 

The scale $f_{S-1}$ cannot exceed 15 GeV in case of the low-scale model. We 
will later explain how this bound is obtained. Consequently, the lightest PGB 
in this model is the color singlet boson associated with symmetry breaking by 
the electron mass and its mass is about $1$ GeV\@. The lightest leptoquarks in 
this model have masses approximately $50$ GeV\@. Such a low-scale model is 
therefore ruled out, as we will show in the next chapter when we discuss 
leptoquark searches. However, in the high-scale models, the scale of the 
$SU(S-1)$ interactions can be much larger. Of course, the same formula holds 
for the contributions to the masses from color and electromagnetic 
interactions. Therefore, the leptoquarks and the octet particles can be quite 
heavy with masses of the order of several hundred GeV's. 

We now estimate the masses of the PGBs in the high-scale models presented 
in Figs.~\ref{fig:highscale} and~\ref{fig:highscaleab}a. We describe 
contributions arising from breaking of the $SU(12)_L$ and $SU(12)_D$ 
symmetries by fermion mass terms. These contributions are different 
from the ones in the low-scale model. The high-scale models do not 
contain multi-fermion operators that could give explicit mass terms for 
the light fermions. Instead, the flavor dynamics is generated at a high 
scale, and its low-energy manifestation is the mixing among the ETC gauge 
bosons in different ETC groups. Such a mixing generates the same masses for 
the ordinary fermions and the light ones.

As mentioned before, we assume that in the model of Fig.~\ref{fig:highscale} 
certain fermion condensates form. The $[12_A,S-1]$ fermions form condensates 
with the $[S-1,n+12_D]$, and $[n+12_L,S-1]$ with $[S-1,12_A]$. In the model
presented in Fig.~\ref{fig:highscaleab}a this assumption is fulfilled 
automatically. As before, we describe the PGBs in terms of their nonlinear
representations $\Sigma_1$ and $\Sigma_2$. $\Sigma_1$ refers to the 
condensate of $[12_A,S-1]$ and $[S-1,n+12_D]$ fermions, $\Sigma_2$
to the condensate of $[n+12_L,S-1]$ and $[S-1,12_A]$.

The $\Sigma_i$ matrices have the following transformation properties
\begin{equation}
  \Sigma_1 \rightarrow A_1^\dagger \Sigma_1 D \; \; {\rm and} \; \;     
  \Sigma_2 \rightarrow L^\dagger \Sigma_2 A_2.
\end{equation}
The gauge group A is weakly gauged, so the two sets of fermion fields
$[12_A,S-1]$ and $[S-1,12_A]$ can transform independently. The 
$SU(12) \times SU(12)$ group of $A_1 \times A_2$ transformations is broken
by small terms proportional to the gauge coupling of $A$\@. The mass term 
for the quarks, leptons and the exotic fermions transforms as
\begin{equation}
  m \rightarrow L^\dagger m D,
\end{equation}
and the generators of the $SU(S-1)_A$ group transform both under the
$A_1$ and $A_2$ matrices:
\begin{equation}
  T_A^a \rightarrow A_1^\dagger T_A^a A_1 \; \; {\rm and} \; \;
  T_A^a \rightarrow A_2^\dagger T_A^a A_2
\end{equation}

\begin{figure}[ht]
  \PSbox{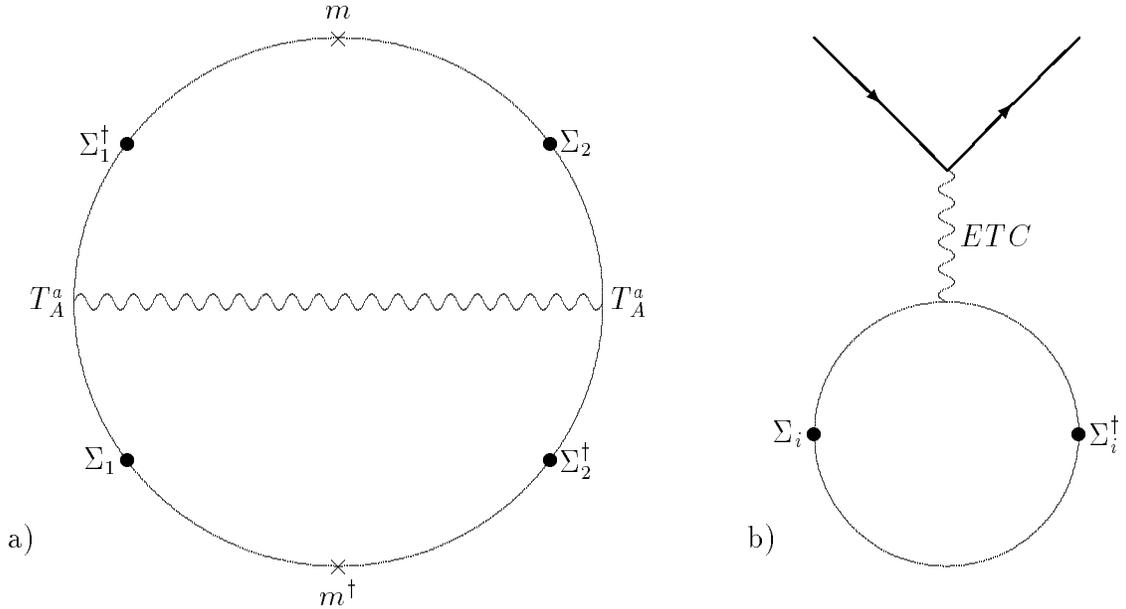 hscale=100 vscale=100 hoffset=-70 
  voffset=-420}{13.7cm}{8.5cm} 
  \caption{a) Diagram contributing to the masses of PGBs in 
              the high-scale models. \newline 
           b) Diagram contributing to the PGB-fermion couplings in the
              high-scale models.}
  \label{fig:mass_coupling}
\end{figure}

Having written all the symmetry properties, we are ready to estimate the 
masses of the PGBs. The lowest-order term contributing to the masses
has the form
\begin{equation}
  \frac{\alpha_{S-1}}{4 \pi} f_{S-1}^2   \, tr\!\left( 
  \Sigma_2 T_A^a \Sigma_2^\dagger m \Sigma_1^\dagger T_A^a 
  \Sigma_1 m^\dagger \right).
\label{eq:highmass}
\end{equation}
Such a contribution arises from the diagram illustrated in 
Fig.~\ref{fig:mass_coupling}a. This term gives masses to the linear
combination of PGBs: $\pi_+^a = \frac{1}{\sqrt{2}}(\pi_1^a + \pi_2^a)$,
where the mass matrix squared is
\begin{equation}
  \Delta m_{ab}^2 = \frac{\alpha_{S-1}}{2 \pi} \, \left[ tr(m)\, 
  tr( T_a T_b m) - tr(T_a m)\,  tr(T_b m) \right].
\end{equation}
The above equation reveals an interesting feature of the PGBs spectrum
in this model. The contribution to masses of the PGBs from the term in
Eq.~\ref{eq:highmass} does not depend on the scale of $SU(S-1)$ interactions.
Thus, the masses of the neutral PGBs do not depend on the scale $f_{S-1}$
in the lowest order. The estimate for the masses depends on the value 
of $\alpha_{S-1}$ and an unknown coefficient of order one. 
For instance, the mass squared of the $P^0$ boson is about $1 \, {\rm GeV}^2$ 
times $\alpha_{S-1}$. Of course, there are also higher-order contributions 
to the PGBs masses. These can arise from the exchange of the ETC gauge 
bosons. Such terms will be proportional to additional powers of 
$\frac{f^2_{S-1}}{f^2_{ETC}}$ which are significant only for a very large
$f_{S-1}$. The masses of color octet and triplet states will 
be much larger due to the gluon exchange contributions described in 
Eq.~\ref{eq:scaleuppi}. The orthogonal combination $\pi_-^a = 
\frac{1}{\sqrt{2}}(\pi_1^a - \pi_2^a)$ remains massless. The $\pi_-^a$ 
bosons are the would-be Goldstone bosons which are eaten when the group
$A$ gets broken. $A$ is completely broken below the scale of the $SU(S-1)$ 
interactions. The only unbroken gauge symmetry at low energies are color 
$SU(3)$ and hypercharge $U(1)$, color group is a linear combination of 
$SU(3)$ subgroups of the $A$ group and the ETC groups. Thus, the 
spectrum consists of 143 PGBs, which is the same number as in 
the low-scale model.

A qualitatively different mass spectrum might be the feature of the high-scale 
model described in Fig.~\ref{fig:highscaleab}b. In this model, the masses of 
the fermions transforming under $SU(S-1)$ are unrelated to the masses of 
quarks and leptons. The masses of the neutral PGBs can arise only from 
explicit mass terms for the $[12_1,S-1_1]$ and $[S-1_1,n+12_D]$ fermions, 
and mass terms for the $[n+12_L,S-1_2]$ and $[S-1_2,12_2]$. Such mass terms 
can result from four-fermion operators created at some very high scale. 

Multi-fermion operators are inevitable in the TC-GIM models. Without their 
presence it is impossible to generate a nontrivial Kobayashi-Maskawa mixing
matrix. A non-diagonal form of the mixing matrix implies that independent 
flavor symmetries for right-handed and left-handed fermions are completely
broken. One cannot achieve such breaking by mass terms only, as such
terms always leave several $U(1)$ symmetries~\cite{CTSM}. The multi-fermion
operators are assumed to be generated at some high scale, perhaps as a result  
of fermion compositeness or some other mechanism. Specific operators needed 
for generating the  Kobayashi-Maskawa matrix are listed in Refs.\ 
\cite{themodel} and \cite{physfromvacuum}. It is plausible that those are
not the only multi-fermion operators present in this model. Additional 
operators may generate masses for the light fermions. For instance,
the mass term for the $[12_1,S-1_1]$ and $[S-1_1,n+12_D]$ fermions
can be generated by the operator 
\begin{displaymath}
  \frac{1}{f_F^2} \, [12_1,S-1_1] [S-1_1,n+12_D] [n+12_D,S_D] [S_D,12_2],
\end{displaymath}
where we have used the notation introduced in Fig.~\ref{fig:model} and 
$f_F$ is the scale at which such operators arise. When $SU(S_D)$ 
interactions become strong, this operator gives a mass term with 
magnitude proportional to $4 \pi f_{S}^3 / f_F^2$.

Although we do not know the particular form of the mass matrix, we can
estimate an upper bound on the magnitude of its elements. The scale of 
$SU(S)$ and $SU(2S)$ interactions is the same as the scale at which the ETC 
groups get broken, since the mechanism that triggers breaking of the ETC 
groups is the dynamical breaking of $SU(S)$ and $SU(2S)$ symmetries. 
Therefore, the exotic fermions have masses not larger than 
\begin{equation}
   \label{eq:maxmass}
   0.04\, {\rm GeV} \, (\frac{f_{ETC}}{1.5 {\rm TeV}})^3 
                    \, (\frac{1000 {\rm TeV}}{f_F})^2.
\end{equation}
The form of the term contributing to the masses of the PGBs is 
identical to the one in Eq.~\ref{eq:mass}; the mass matrix should be 
replaced with the mass matrix for the relevant exotic fermions. As a result,
the contributions to the masses cannot exceed approximately 
$8 \,{\rm GeV} \, \sqrt{\frac{f_{S-1}}{15 \,{\rm GeV}}}$. For the neutral 
PGBs this is the only contribution, and it can be treated as an upper bound
on their masses. Of course, in order to obtain masses of colored PGBs, one 
should add independent contributions to the square of PGBs masses. In this 
case, the above contribution from explicit breaking terms has to be added 
to the gluon exchange contribution described in Eq.~\ref{eq:scaleuppi}.

We now describe the couplings of the PGBs to the quarks and leptons.
The symmetry properties of the quarks, leptons and the low-scale PGBs 
allow very simple invariant couplings:
\begin{equation}
  \label{eq:lowcoupling}
  \frac{4 \pi f_{S-1}^3 v_{TC}^2}{f_{ETC}^4} \left( \bar{\psi_L} \Sigma
  d_R + \ {\rm h.c.} \right),
\end{equation}
where the coefficient $\frac{v_{TC}^2}{f_{ETC}^4}$ reflects the fact that 
such operator is created by ETC interactions. The lowest-dimension term in 
the above equation is the contribution of the $S-1$ condensate to the masses 
of the ordinary fermions. Such a contribution cannot be larger than the 
electron mass, which limits the scale $f_{S-1}$ to be less than
$15 \ {\rm GeV}$. Given this constraint, the masses of PGBs are small, 
such that the Tevatron experiments should see signals of leptoquarks.
We want to make leptoquarks as heavy as possible, saturating the bound,
and then see if such PGBs can evade detection. Therefore, we assume that
the factor $\frac{4 \pi f_{S-1}^3 v_{TC}^2}{f_{ETC}^4}$ is numerically 
equal to the electron mass.

The PGBs in the high-scale models exhibit dramatically different
couplings to the quarks and leptons. Such couplings are generated
by diagrams of the type presented in Fig.~\ref{fig:mass_coupling}b.
The relevant terms are:
\begin{equation}  
  \label{eq:highcoupling}
  \left( \frac{f_{S-1}}{f_{ETC}} \right)^2
  \bar{\psi_L} \gamma^\mu (\partial_\mu \Sigma_2) \Sigma_2^\dagger \psi_L
  \; {\rm and} \;
  \left( \frac{f_{S-1}}{f_{ETC}} \right)^2
  \bar{d_R} \gamma^\mu (\partial_\mu \Sigma_1) \Sigma_1^\dagger d_R
\end{equation}
Thus, the PGBs couple derivatively to the quarks and leptons. We should 
mention here that these couplings are quite different from couplings 
predicted for PGBs in other technicolor models~\cite{walkingTC,variousTC}.
In generic TC models, the couplings are non-derivative and proportional 
to the fermion masses. Typically, they are of the form
\begin{displaymath}
\frac{m_q}{f_\Pi} \, \Pi \, \bar{q} \gamma_5 q.
\end{displaymath}
Such terms have a large magnitude for the coupling to the top quark.
In TC-GIM models the magnitudes of couplings are much smaller and are 
also suppressed by the  momentum of PGBs. This difference in couplings 
invalidates bounds on PGBs and leptoquarks obtained by the studies of 
flavor-changing neutral currents mediated by those 
particles~\cite{Leurer,technipionbsg}.

The PGBs also couple to gauge bosons. We describe both the minimal couplings
of charged or colored particles and couplings arising from the anomalies.
The constituents of the PGBs are singlets of the weak $SU(2)$ interactions,
thus the PGBs do not couple to the $W^\pm$ bosons. Couplings to the photon
and the $Z^0$ originate from the hypercharge interactions. The lowest-order
couplings are contained in the kinetic energy term for the PGBs
\begin{equation}
\label{eq:vectorcouplings}
 \begin{array}{c}
  L_{kin}=
  \frac{f_{S-1}^4}{4} tr\left[(D^\mu \Sigma)^\dagger (D_\mu \Sigma)\right],
  \\
  D_\mu \Sigma = \partial_\mu \Sigma + i e A_\mu [Q,\Sigma] + 
  i e \tan(\theta_W) Z_\mu [Q,\Sigma] + i g_3 G^a_\mu [\lambda_a,\Sigma],
 \end{array}
\end{equation}
where $\theta_W$ is the Weinberg angle and $Q$ is the fermion charge 
matrix. Only the leptoquarks couple to the photon and the $Z^0$, while 
other PGBs are neutral.

The anomalous couplings are restricted to particles that are singlets with
respect to flavor symmetries. Thus, the $P^0$ is the only particle
that couples to a photon pair. Color octet bosons $\theta_a$ couple
both to a pair of gluons and to a photon and a gluon. A general coupling
of a PGB to a pair of gauge bosons can be written \cite{anomalous} as
\begin{equation}
  \label{eq:anomgeneral}
  (S-1) A \frac{g_1 g_2}{\pi^2 f_{S-1}} \epsilon_{\mu \nu \rho \sigma}
  k_1^\mu k_2^\nu \epsilon_1^\rho \epsilon_2^\sigma,
\end{equation}
where $k_i$ and $\epsilon_i$ are the momenta and polarizations of the 
gauge bosons, $g_i$ are the coupling constants. Below, we list all the 
coefficients of non-vanishing anomalous couplings:
\begin{equation}
  \label{eq:anompart}
  \begin{array}{cc}
  A_{P^0 \gamma \gamma} = \frac{(-2,1)}{3 \sqrt{2}} &
  A_{P^0 g_a g_b} = \frac{1}{16 \sqrt{2}} \delta_{ab} \\
  A_{\theta_a g_b g_c} = \frac{\sqrt{3}}{8} d_{abc}   &
  A_{\theta_a g_b \gamma} = \frac{(-1,2)}{4\sqrt{3}} \delta_{ab} \, .
  \end{array}
\end{equation}
The coefficient $d_{abc}$ is the $SU(3)$ symmetric structure constant.
The constants $A_{P^0 \gamma \gamma}$ and $A_{\theta_a g_b \gamma}$
are different for the up and down sectors because the charge matrices
in those sectors are not identical. The first number in parenthesis
refers to the down sectors and the second one to the up sector. Obviously, 
the $Z^0$ coupling has the same structure as the photon coupling. Therefore,
every photon field in the above equations can be replaced by $Z^0$,
while the electric charge is replaced by $e \tan\theta_W$.

\section{Decays, production rates and signatures}
\label{sec:rates}

In this section we describe decays of the PGBs and discuss 
various production mechanisms at both electron and hadron colliders.
We present predictions for the operating machines -- LEP, HERA and Tevatron,
and also for the planned colliders: upgraded Tevatron, LHC and NLC.
The section ends with a comparison of PGBs' features in TC-GIM models with
other technicolor scenarios. 

The PGBs decay dominantly into fermion-antifermion pairs, but the decay widths
are model dependent. In the low-scale model, decays are governed by the
couplings described in Eq.~\ref{eq:lowcoupling} and the corresponding 
decay widths are
\begin{eqnarray}
  \Gamma_{low}(\Pi^a\rightarrow f_i \bar{f_j}) & = &
  \frac{2 m_e^2 |T^{\Pi^a}_{ij}|^2}{\pi M_{\Pi^a}^2 f_{S-1}^2} k 
  \left( \sqrt{k^2+m_i^2} \sqrt{k^2+m_j^2} + k^2 + m_i m_j 
  \right)      
  \\
  & \approx & \left\{ \begin{array}{ll} 
  \frac{m_e^2 |T^{\Pi^a}_{ij}|^2}{2 \pi} \frac{ M}{f_{S-1}^2} 
  \sqrt{1-\frac{4 m^2}{M^2}} & {\rm for}\ m_i \approx m_j = m  \\
  \frac{m_e^2 |T^{\Pi^a}_{ij}|^2}{2 \pi} \frac{ M}{f_{S-1}^2}
  (1-\frac{m^2}{M^2})^2  & {\rm for}\  m_i \ll m_j =m             
  \end{array} \right., \nonumber 
\end{eqnarray}
where $k^2=\frac{(M-m_i-m_j)(M-m_i+m_j)(M+m_i-m_j)(M+m_i+m_j)}{4 M^2}$,
and $T^{\Pi^a}$ are $SU(12)$ matrices described in Eq.~\ref{eq:PGBs}.
These decay widths do not depend, except for the kinematical factors,
on fermion masses. For decays into two fermions much lighter than a $\Pi^a$,
the decay width equals approximately $0.18 \, {\rm eV}\, |T^{\Pi^a}_{ij}|^2 
\, (\frac{15\, {\rm GeV}}{f_{S-1}})^2 \, \frac{M_{\Pi^a}}{1\, {\rm{GeV}}} $.
The scale of the $SU(S-1)$ interactions is restricted in low-scale model
to be less than $15\, {\rm GeV}$, so all PGBs are very short-lived and decay 
inside a detector.

The high-scale PGBs couple derivatively to the quarks and leptons, as
described in Eq.~\ref{eq:highcoupling}. The resulting decay widths are 
\begin{eqnarray}
  \label{eq:ffdecays}
  \Gamma_{high}(\Pi^a\rightarrow f_i \bar{f_j}) & = & 
  \frac{f_{S-1}^2}{f_{ETC}^4} \, 
  \frac{|T^{\Pi^a}_{ij}|^2 k}{2 \pi M_{\Pi^a}^2}
  [(m_i^2+m_j^2) (M_{\Pi^a}^2-m_i^2-m_j^2) + 4 m_i^2 m_j^2) \\
  & \approx & \left\{ \begin{array}{ll}
  \frac{f_{S-1}^2}{f_{ETC}^4} \, \frac{|T^{\Pi^a}_{ij}|^2}{2 \pi}
  m^2 M \sqrt{1-\frac{4 m^2}{M^2}} & {\rm for}\ m_i\approx m_j = m \\
  \frac{f_{S-1}^2}{f_{ETC}^4} \, \frac{|T^{\Pi^a}_{ij}|^2}{4 \pi}
  m^2 M (1-\frac{m^2}{M^2})^2 & {\rm for}\  m_i \ll m_j =m
  \end{array} \right. . \nonumber
\end{eqnarray}
These decay widths are proportional to the masses squared of the fermions 
in the final state. This is a result of  chiral suppression, similar to 
the familiar $\pi^+\rightarrow \mu^+ \nu_\mu$  decay. PGBs decays into
fermions are caused by the ETC interactions, which preserve lepton and
quark flavor. Therefore, the PGBs decay into quarks and leptons of the 
same flavor as PGBs constituents. The partial width of the $P^0$ decay
into a $\mu^+\mu^-$ pair equals approximately $1.5 \cdot 10^{-5} \,{\rm eV}
\frac{M}{1\,{\rm GeV}}(\frac{f_{S-1}}{15\,{\rm GeV}})^2
(\frac{1.5\,{\rm TeV}}{f_{ETC}})^4$. We frequently use $P^0$ as an 
illustration because this particle couples to pairs of photons and gluons, 
and therefore its properties will be very important later on. It is very 
interesting that PGBs in the high-scale models are also very narrow 
resonances, even the heaviest scalars have widths smaller than 1 GeV.

Few PGBs have anomalous couplings to gluons and photons. The couplings
are described in Eqs.~\ref{eq:anomgeneral} and \ref{eq:anompart}.
The resulting widths of PGBs decays into a pair of massless vector 
bosons are
\begin{equation}
  \label{eq:vvdecays}
  \Gamma(\Pi^a\rightarrow V_i V_j ) = | A_{\Pi^a V_i V_j} |^2 \,
  \frac{g_i^2 g_j^2}{32 \pi^5} \, (S-1)^2 \, \frac{M^3}{f_{S-1}^2} \,
  \frac{1}{1+\delta_{V_i V_j}} \, .
  \end{equation}
For example, the width of the decay $P^0 \rightarrow \gamma \gamma$
equals $ 0.16 \, {\rm eV} \, (S-1)^2 \, (\frac{M}{1 \, {\rm GeV}})^3
\, (\frac{15 \, {\rm GeV}}{f_{S-1}})^2$.

We present a summary of the particle spectrum of the high-scale models in 
Table 1. We list the dimension of $SU(3)_C$ representation the PGBs belong 
to, their masses and major decay modes. Which decay modes dominate depends on 
the scale of $SU(S-1)$ interactions and particle masses. The heavier the 
particle the more important the decays into vector bosons are, since their 
widths grow with the mass cubed, while the widths of fermionic decays are 
linearly proportional to PGBs masses. The scale $f_{S-1}$ suppresses 
decays into vector bosons, while it enhances fermionic decay modes,
compare Eqs.~\ref{eq:ffdecays} and \ref{eq:vvdecays}.
\begin{table}
  \begin{center}
  \begin{tabular}{|c|c|c|l|} \hline
    particle    & $SU(3)$ & mass [GeV]  & decay modes              \\ \hline
    $\theta^i_a$& 8       & 45 ($f/15$) & $q \bar{q}$              \\ \hline
    $\theta_a$  & 8       & 45 ($f/15$) & $gg, \gamma g, q\bar{q}$ \\ \hline
    $T^i_c, T_c$& 3       & 31 ($f/15$) & $q \bar{l}$              \\ \hline
    $\Pi^i, P^i$& 1       & 0.1--5      & $q \bar{q}, l \bar{l}$   \\ \hline
    $P^0$       & 1       & 1           & $gg, q\bar{q}, l\bar{l}$ \\ \hline
  \end{tabular}
  \end{center}
  \caption{The particle spectrum in the high-scale models.}
\end{table}   

\subsection{Electron Colliders}
Charged PGBs can be pair-produced in $e^+ e^-$ collisions. Charged
bosons, which are only leptoquarks, couple both to the photon and
to the $Z^0$ with coupling described in Eq.~\ref{eq:vectorcouplings}.
Thus, the leptoquarks can be produced at the $Z^0$ peak and also at energies
above the $Z^0$ mass, where both photon and $Z^0$ exchanges contribute
to the production rate. The decay width of a $Z^0$ into a pair of 
leptoquarks is
\begin{equation}
  \Gamma(Z^0 \rightarrow T \bar{T}) = \alpha \,
  \left( \frac{2}{3} \tan\theta_W\right)^2
  \, \frac{M_Z}{4} \left(1-\frac{4 m_T^2}{M_Z^2}\right)^{\frac{3}{2}}.
\end{equation}
Signatures of such events are quite distinct, each leptoquark decays into
a hadronic jet and an isolated lepton. The events would have two 
opposite-sign leptons, two hadronic jets and no missing energy.
The main background comes from the $Z^0 \rightarrow b \bar{b}$ decays,
followed by semileptonic decays of both $b$ quarks. However, the event shape
is different and the background can be efficiently rejected.  
Searches performed at LEP exclude pair-produced leptoquarks up to
$45.5 \, {\rm GeV}$~\cite{DELPHI}, which is almost the kinematic 
limit. The limit reported in Ref.~\cite{DELPHI} does apply to 
leptoquarks in our model, even though both the coupling to the $Z^0$  
and decay modes differ. The couplings of the $Z^0$ to a pair of leptoquarks
in our model and leptoquarks in superstring-inspired models~\cite{DELPHI} 
result in cross sections that are numerically very close. Leptoquarks in our 
model have larger charge, but they do not interact weakly, and the two effects
roughly compensate. The decay mode $T \rightarrow d e^+$ is experimentally 
indistinguishable from the mode $T \rightarrow \bar{u} e^+$, which was 
searched for at LEP.

There are two production mechanisms of single PGBs at the $Z^0$ pole.
A $Z^0$ can decay into a fermion pair, and subsequently a PGB is radiated
off a fermion line. Such a mechanism is model dependent, since the magnitudes
of couplings of PGBs to fermions are not dictated by the gauge invariance.
Corresponding decay widths in the low-scale model are
\begin{equation}
  \begin{array}{l}
  \Gamma_{low}(Z^0\rightarrow f_i \bar{f_j} \Pi^a) = \\
  \; \; \left( \frac{g \, m_e}{\cos \theta_W f_{S-1}}\right)^2
  | T_{ij}^{Pi^a} |^2 \, \frac{(g_V^2 +g_A^2) M_Z}{576 \pi^3} \, 
  (-17+9 r +9r^2 - r^3 -6\log r - 18 r \log r),
  \end{array}
\end{equation}
where $r=(\frac{m_{\Pi^a}}{M_Z})^2$. Meanwhile, 
$g_V=\frac{1}{2}T_3-Q \sin^2 \theta_W$ and $g_A=\frac{1}{2} T_3$ are 
vector and axial couplings of the $Z^0$ to an $f_i \bar{f_i}$ pair. 
We assumed that fermion masses are negligible compared to the $Z^0$ mass. 
The decay rate diverges as the mass of the scalar particle approaches zero, 
which is a result of divergent fermion propagator when a light scalar is 
being emitted collinearly with the fermion. However, for reasonable 
values of PGB masses, even as light as 1 GeV, this decay width is orders 
of magnitude too small for such a process to be observed. 

PGBs couplings to fermions are different in the high-scale models, so
the decay width has a different form
\begin{equation}
  \begin{array}{l}
  \Gamma_{high}(Z^0\rightarrow f_i \bar{f_j} \Pi^a) = \\
  \; \; \left( \frac{g f_{S-1} M_Z}{\cos \theta_W f^2_{ETC}} \right)^2
  | T_{ij}^{Pi^a} |^2 \, \frac{(g_V \mp g_A)^2 M_Z}{1152 \pi^3} \, 
  (1+9 r -9r^2 -r^3 +6 r \log r +6 r^2 \log r),
  \end{array}
\end{equation}
where $g_V - g_A$ applies to the PGBs that couple to right-handed fermions,
and $g_V+g_A$ to left handed ones. PGBs in the model described in 
Figs.~\ref{fig:highscale} and \ref{fig:highscaleab}a are linear combinations 
of both types of PGBs, they couple to the left and right-handed fermions. 
This decay width is not divergent for small masses of the scalar, the 
coupling of the scalar is proportional to its momentum, which annihilates 
divergence of the fermion propagator. For light PGBs the decay width can 
be approximated as $\Gamma(Z^0\rightarrow f_i \bar{f_j} \Pi^a) 
\approx 0.13 \cdot 10^{-7} {\rm GeV} \; (\frac{f_{S-1}}{150\;{\rm GeV}})^2 
\; (\frac{1.5 \;{\rm TeV}}{f_{ETC}})^4 \; | T_{ij}^{Pi^a} |^2 $, which is 
too small to be observed at LEP. In principle, this process could provide 
an upper bound on the scale $f_{S-1}$. If the scale $f_{S-1}$ is very large, 
of the order of 1 TeV, and at the same time the scale of the ETC interactions 
is also around 1 TeV, some events could be observed at LEP\@. However, such 
a case is not too interesting, because the hierarchy of symmetry breaking 
in the model would not work as expected. Such a small decay width of the 
$Z^0$ into a PGB and a fermion pair makes this process impossible to observe
at LEP. Mass limits on singly-produced leptoquarks presented in 
Ref.~\cite{DELPHI} assume a different form of PGBs' couplings to 
fermions. Therefore, those limits are not valid in the TC-GIM models.

Other sources of PGBs production are anomalous couplings to a $Z^0$ and
a photon. The width of the $Z^0$ decay into a PGB and a photon has been 
calculated in Ref.~\cite{PGBatZpole}. We use their results together 
with anomaly factors from Eq.~\ref{eq:anompart} and obtain the decay width
\begin{eqnarray}
  \label{eq:ztopgamma}
  \Gamma(Z^0 \rightarrow P^0 \gamma) & = &
  \left( \frac{A_{\gamma\gamma}^{P^0} \; (S-1)}{f_{S-1}} \right)^2 
  \frac{\alpha^2 \tan^2 \theta_W}{6 \pi^3} \; M_Z^3 
  \left( 1 -\frac{m_{P^0}^2}{M_Z^2} \right)^3 \\
  & = & 1.9 \; 10^{-5} {\rm GeV} \; (S-1)^2 \; 
  \left(\frac{15 {\rm GeV}}{f_{S-1}}\right)^2
  \left( 1 -\frac{m_{P^0}^2}{M_Z^2} \right)^3 \ \times (4,1), \nonumber
\end{eqnarray}
where 4 refers to the down-type $P^0$ and 1 to the up-type. This decay 
rate is large enough to constrain the scale of the $SU(S-1)$ interactions.
For numerical estimates we will assume that $S=4$. The signature of such 
events is quite unique -- an isolated monoenergetic photon and $P^0$ decay 
products. A one GeV $P^0$ boson in the high-scale models decays most likely 
into a small number of pions or a pair of $K$ mesons, therefore the events
are characterized by very low hadron multiplicity. Such a signature is 
similar to the signature of the $Z^0 \rightarrow \eta ' \gamma$ decays,
where the $\eta '$ decay products contain a $\pi^+\pi^-$ pair. The ALEPH
Collaboration reported that the $BR(Z^0 \rightarrow \eta ' \gamma)$ is less
than $4.2 \cdot 10^{-5}$\cite{ALEPH}. Assuming that the branching ratio of 
the decays $P^0 \rightarrow \pi^+ \pi^- X$ is at least 50\%, same bound can
be placed on the branching ratio of the $Z^0 \rightarrow P^0 \gamma$ decays.
Using Eq.~\ref{eq:ztopgamma} this branching ratio translates to a lower limit
of 38 GeV on $f_{S-1}$. At the time of the analysis the ALEPH collaboration 
collected only $8.5 \, {\rm pb^{-1}}$ of integrated luminosity. Currently, 
the LEP experiments have data from over $100\, {\rm pb^{-1}}$. If all the
data were analyzed one could place a limit of $10^{-5}$ on the 
$BR(Z^0 \rightarrow P^0 \gamma)$. Such a limit corresponds to exploring the 
scale $f_{S-1}$ up to 80 GeV\@. For the high-scale model depicted in 
Fig.~\ref{fig:highscaleab}b, we obtain a similar number. Even 
if $P^0$ is as heavy as it could possibly be, given the estimate in 
Eq.~\ref{eq:maxmass}, the lower limit on $f_{S-1}$ is 75 GeV\@. The bound
of 80 GeV would surpass the results from the Tevatron, where experiments
currently probe $f_{S-1}$ up to 65 GeV.

We now turn our attention to future $e^+e^-$ colliders --- LEP2 operating 
above the $W^+W^-$ threshold and a collider with the CM energy of 500 GeV, 
which is one of the options for the proposed Next Linear Collider. We assume
that LEP2 will collect $500 \, {\rm pb^{-1}}$ of integrated luminosity per
year of running~\cite{LEP2} and the NLC will achieve its design luminosity of
$50\, {\rm fb^{-1}}$ per year~\cite{NLC1,NLC2}.
 
The most obvious process to look for is pair production of leptoquarks.
The cross section for this process is
\begin{equation}
  \label{eq:eett}
  \begin{array}{l}
  \sigma(e^+e^- \rightarrow T\bar{T}) = \xi \, \left(\frac{2}{3}\right)^2 \,
  \frac{\pi \alpha^2}{s} \left( 1- \frac{4 m^2}{s} \right)^{3/2}, \\
  \xi=\left(1+\frac{g_V}{\cos^2 \theta_W (1-y)} \right)^2
  + \left(\frac{g_A}{\cos^2 \theta_W (1-y)} \right)^2,
  \end{array}
\end{equation}
where $\frac{2}{3}$ is the leptoquark charge, $y=\frac{m_Z^2}{s}$,
$g_V$ and $g_A$ are the vector and axial couplings of the $Z^0$ to an
$e^+e^-$ pair. The reaction is mediated by both photon and $Z^0$ exchanges,
which are both included in the derivation of Eq.~\ref{eq:eett} and manifest 
as the factor $\xi$. Such a simple result for the two diagrams and their 
interference is caused by the simplicity of the $Z^0$ couplings to the exotic 
particles. The $Z^0$ couples exactly the way photon does with the coupling 
multiplied by $\tan \theta_W$. The signatures of such events are relatively 
easy to disentangle from the backgrounds. Therefore, leptoquarks can be 
discovered if their masses are only few GeV smaller than the kinematic limits.
The potential for leptoquark discovery can be translated into limits on the 
scale $f_{S-1}$ using the leptoquark mass estimate from 
Eq.~\ref{eq:scaleuppi}. LEP2 can probe the scale $f_{S-1}$ up to 45 GeV 
while NLC up to 120 GeV.

Single PGB production via anomaly coupling is a process whose importance
grows with energy. The anomaly coupling of a PGB and two vector bosons
is proportional to vector boson momentum, so the production cross section
does not decrease with $\sqrt{s}$. The cross section for producing
a $P^0$ and a photon equals 
\begin{eqnarray}
\label{eq:eetopgamma}
  \sigma(e^+e^- \rightarrow \gamma P^0) &  = & \xi \, 
  \frac{2 \alpha^3 (S-1)^2}{3 \pi^2 f_{S-1}^2}
  (A_{\gamma \gamma}^{P^0})^2 \left(1-\frac{m^2_{P^0}}{s} \right)^3 \\
  & \approx & 3.7\, {\rm fb} \, (S-1)^2 
  \left(\frac{15\,{\rm GeV}}{f_{S-1}}\right)^2
  \left(1-\frac{m^2_{P^0}}{s} \right)^3 \ \times (4,1), \nonumber
\end{eqnarray}
where, again, 4 refers to the down-type $P^0$ and 1 to the up-type one. 
The study of this process at LEP2 will not provide any new information
beyond what we already know from the $Z^0$ decays, the luminosity will be too 
small to produce any events. If $f_{S-1}$ is smaller than about 200 GeV
dominant decay modes of $P^0$ are hadronic. For larger values of $f_{S-1}$
the $P^0 \rightarrow \mu^+\mu^-$ will dominate. This is a great help in the
detection of $P^0$ as $\mu^+\mu^-$ pairs are measured with large efficiency 
and good angular resolution. The signature is then a monoenergetic photon and 
a $\mu^+\mu^-$ pair with the invariant mass about 1 GeV\@. Such events have 
very little background, so we assume that as few as 10 produced $P^0$ bosons 
are enough to be detected. Consequently, the NLC is likely to probe the scale 
of $SU(S-1)$ interactions up to 390 GeV\@. As before, limits on the $f_{S-1}$ 
scale in the model depicted in Fig.~\ref{fig:highscaleab}b are at most few 
GeV lower than the limits in other high-scale models. One can also look for 
$P^0$ produced together with a $Z^0$. The corresponding cross section
\begin{equation} 
  \sigma(e^+e^- \rightarrow Z^0 P^0) = \tan^2\theta_W \xi \, 
  \frac{2 \alpha^3 (S-1)^2}{3 \pi^2 f_{S-1}^2}
  (A_{\gamma \gamma}^{P^0})^2 
  (1-\frac{(m_Z+m_{P^0})^2}{s} )^\frac{3}{2}
  (1-\frac{(m_Z-m_{P^0})^2}{s} )^\frac{3}{2}
\end{equation}
is about 28\% of that for $e^+e^- \rightarrow \gamma P^0$. This process 
can be useful only for relatively small scales $f_{S-1}$, not larger than 
100 GeV\@. A large number of events is needed because the $Z^0$ can be 
measured precisely only in leptonic channels, whose branching ratios are 
small.

\subsection{Hadron Colliders}
There are variety of processes in which the PGBs can be produced in hadron
colliders. Gluon-gluon and quark anti-quark annihilations are sources of
PGB pair production. Quark-gluon fusion produces single PGBs. The production
of single PGBs via anomalous couplings to two gluons is also a possibility.
Unfortunately, neutral PGBs, with the exception of $P^0$, do not have large
enough production rates to be observed in hadron collisions. PGBs' couplings
to fermions are too small to give significant cross section. For this reason,
HERA does not provide any information about PGBs in the TC-GIM models.

\begin{figure}[h]
  \PSbox{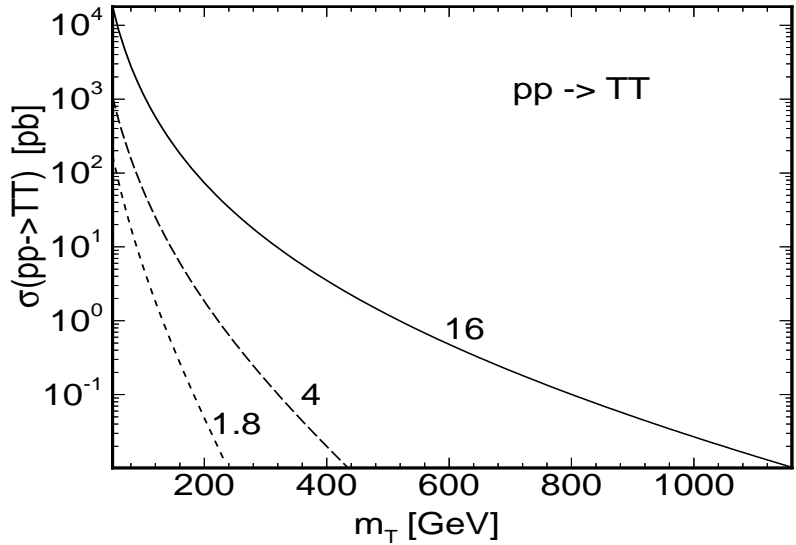 hscale=70 vscale=58 hoffset=20 
         voffset=-190}{13.7cm}{7cm} 
  \caption{The cross section for pair production of leptoquarks in pp 
           collisions. The three curves correspond to $\protect\sqrt{s}=$
           1.8, 4 and 16 TeV.}
  \label{fig:lq}
\end{figure}

\begin{figure}[!t]
  \PSbox{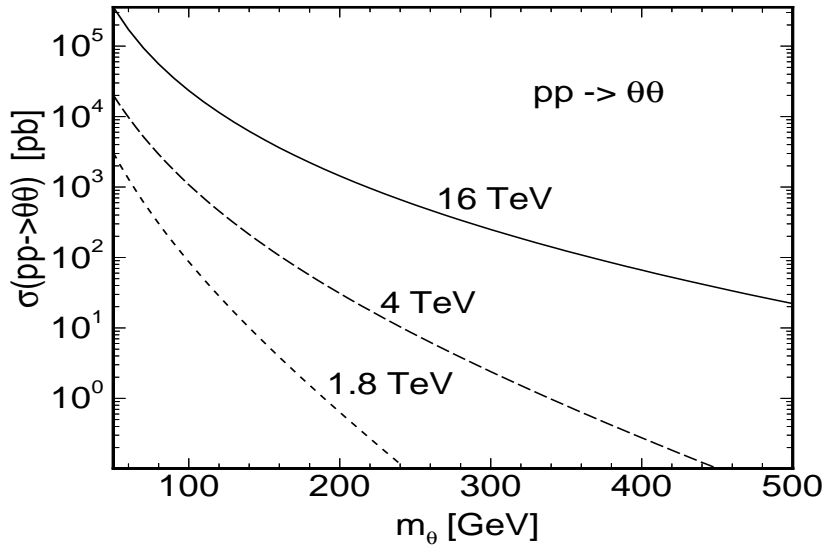 hscale=70 vscale=58 hoffset=20 
  voffset=-190}{13.7cm}{7cm} 
  \caption{The cross section for pair production of color-octet PGBs in pp
           collisions. Different curves correspond to the CM energy at the
           Tevatron, upgraded Tevatron and the LHC.}
  \label{fig:octet}
\end{figure}

The cross section for the pair production of PGBs has been calculated 
in Ref.~\cite{EHLQ} for the general case of scalar particles in the 
D-dimensional representation of color $SU(3)$. The quark anti-quark
fusion cross section is
\begin{equation}
  \frac{d \sigma}{d \hat{t}}(q \bar{q} \rightarrow \Pi \Pi) = 
  \frac{2 \pi \alpha_s^2}{9 \hat{s}^2} \, k_D \beta^2 (1-z^2)
\end{equation}
and for gluon-gluon annihilation
\begin{equation}
  \frac{d \sigma}{d \hat{t}}( g g  \rightarrow \Pi \Pi) =
  \frac{2 \pi \alpha_s^2}{\hat{s}^2} \, k_D 
  \left( \frac{k_D}{D} -\frac{3}{32} (1-\beta^2 z^2) \right)
  \left( 1 -2 V + 2 V^2 \right) .
\end{equation}
In the above formulas, $k_D$ is the Dynkin index of the D-dimensional 
representation ($k_3=\frac{1}{2}$, $k_8=3$), $z$ is the cosine of parton 
scattering angle in the center of mass, 
\begin{displaymath}
V=1-\frac{1-\beta^2}{1-\beta^2 z^2} \ \ {\rm and} \ \ 
\beta^2=1-\frac{4 m_\Pi^2}{\hat{s}},
\end{displaymath}
while $\hat{s}$ and $\hat{t}$ are Mandelstam variables for the annihilating
partons. Using these formulas and parton distributions from 
Ref.~\cite{DukeOwens} (set 1), we obtain production rates for 
leptoquarks and octet particles. The cross sections are presented in 
Figs.~\ref{fig:lq} and \ref{fig:octet}, which agree with the results 
of Refs.~\cite{EHLQ,HewettPakvasa}. 

Leptoquarks are a feature of many extensions of the Standard 
Model~\cite{otherleptoquarks}. Several conclusions about leptoquark
searches do apply to TC-GIM models. For instance, pair production of
leptoquarks is almost model independent. Since gluon-gluon annihilation 
dominates over quark anti-quark annihilation, the production rates do not 
depend on leptoquarks' couplings to quark pairs. There is a difference, 
however: in most models leptoquarks decay into a quark and a lepton of
the same generation. In the TC-GIM models, the leptoquarks carry lepton 
and quark numbers of any generation. There are single-generation leptoquarks,
but there exist leptoquarks that mix different generations as well.

A leptoquark decaying into an electron and a d quark has the same signature 
as the so-called first generation leptoquark. The first generation 
leptoquarks decay into an electron and a first generation quark with
branching ratio $\beta$ or into a neutrino and a quark with branching ratio
$1-\beta$. The experimental limits depend on the unknown ratio $\beta$.
The down-type leptoquarks in the TC-GIM models have $\beta=1$, while
the up-type ones have $\beta=0$.  The strongest limit on the leptoquark masses
has been obtained by the D0 Collaboration~\cite{D0}. Their results exclude
down-type leptoquarks up to 130 GeV\@. The second generation leptoquarks 
are excluded up to 133 GeV~\cite{CDF}. These results limit the scale
$f_{S-1}$ to be larger than approximately 65 GeV\@. That is why, as we
previously claimed, the low-scale model is excluded. 

The signatures of pair-produced leptoquarks are quite unique. Hundred 
$T\bar{T}$ pairs should suffice to discover leptoquarks at the 
$\sqrt{s}=4 \, {\rm TeV}$ upgraded Tevatron. Using the cross sections from 
Fig.~\ref{fig:lq}, we estimate that the upgraded Tevatron can push the 
leptoquark mass limit up to 440 GeV, when $10\, {\rm fb}^{-1}$ is collected 
in one year of running~\cite{Tevatronupgrade}. At the LHC, one expects CM 
energy of 16 TeV and the integrated luminosity of $100\, {\rm fb}^{-1}$ per 
one year of running~\cite{LHC}, so the LHC can discover leptoquarks up to 
approximately 1160 GeV\@. Thus, the $f_{S-1}$ scale can be probed up to 215 
GeV at the upgraded Tevatron and  up to 560 GeV at the LHC.  

We now turn our attention to color-octet particles. The production cross
sections for octets are about an order of magnitude larger than those
for leptoquarks due to color factors. Unfortunately, the detection of 
octet particles is difficult. Octet PGBs decay into two hadronic jets, 
so pair-produced octets yield four-jet signals. The QCD four-jet production
is the main source of background, the QCD resulting rate has been 
estimated~\cite{jetbackground}, and it is quite large. The authors of 
Ref.~\cite{multijet} have studied four-jet processes as a probe
of new physics signals. They propose certain kinematic variables designed 
to study such events. First, out of the three possible groupings
of four jets into pairs the one that gives most equal invariant masses
is chosen. The average of the two invariant masses, called balanced doublet
mass, is an important parameter in the study. Then, a strong cut on
the transverse jet momentum is imposed. The value of the transverse momentum 
cut depends on the mass of the particle one looks for. The QCD background
peaks at approximately $3 p_T^{min}$, so particles lighter than $3 p_T^{min}$
can be observed by using such a cut. An excess of events on the 
balanced-doublet mass plot would be a signal of pair-produced octet 
particles. The authors conclude that a 375 GeV octet PGBs can be detected 
at the LHC.

This is a rather modest  discovery potential given the fact that octet 
particles are about $1.5$ times heavier than the leptoquarks.  TC-GIM
models have 18 different color-octet particles, nine in each sector.
If these octet particles are nearly degenerate in mass they may give a 
stronger signal. However, if the mass splittings are comparable to the 
invariant mass resolution of four-jet signals, the situation might be 
more complicated since a wider peak is more difficult to disentangle
from the background. The search for the octet particles will be interesting
only if the upgraded Tevatron discovers leptoquarks lighter than 250 GeV\@.
Then, one can expect to see signals of color octet particles at the LHC.

\begin{figure}[t]
  \PSbox{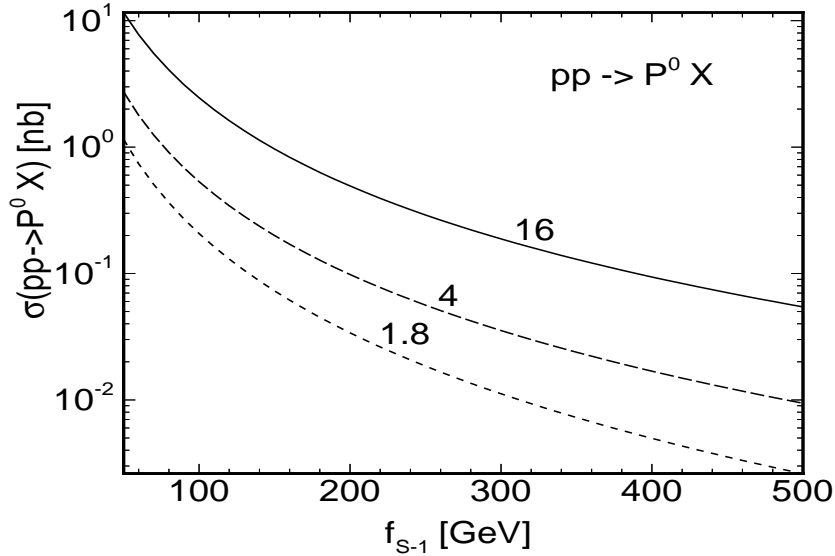 hscale=70 vscale=58 hoffset=20 
  voffset=-180}{13.7cm}{7.5cm} 
  \caption{The cross section for $P^0$ production in pp collisions as a
           function of the $f_{S-1}$ scale. The curves correspond to 
           $\protect\sqrt{s}=$ 1.8, 4 and 16 TeV.}
  \label{fig:anomalous}
\end{figure}

There are several sources of single PGB production in hadron colliders.
A process that gives quite a large ratio is gluon-gluon annihilation
into $P^0$. We compute the production cross section using the narrow
width approximation:
\begin{equation}
  \frac{d \sigma}{d y}(p p \rightarrow P^0 X) =
  \frac{\pi^2}{8 s} \, \frac{\Gamma(P^0\rightarrow g g)}{m_{P^0}} \,
  f_g(\sqrt{\tau} e^y) f_g(\sqrt{\tau} e^{-y}),
\end{equation}
where $\tau=\frac{m_P^2}{s}$. The cross sections of $P^0$ productions are 
depicted in Fig.~\ref{fig:anomalous}. Since the width of the 
$P^0\rightarrow g g$ decay is inversely proportional to $f_{S-1}^2$,
so is the production rate. Despite the fact that the cross section is quite
large, a light $P^0$ cannot be detected in hadron colliders. A one GeV $P^0$
decays predominantly into pions. Such a process does not stand out from
QCD background. The branching ratio for the $P^0\rightarrow \gamma \gamma$
decay is about 2\%, which still does not help much. A one GeV $P^0$ would
decay into two almost collinear photons, which cannot be distinguished.    
Such a light $P^0$ cannot be observed at high energy hadron colliders.

PGBs can also be produced by the quark-gluon fusion. However, the relevant
Feynman diagrams always involve PGBs coupling to fermion pairs. Such
couplings are too small to give significant cross sections.

\subsection{PGBs in TC-GIM versus other technicolor models}
In this subsection we summarize the properties of PGBs in the high-scale 
models presented in Figs.~\ref{fig:highscale} and \ref{fig:highscaleab}a. 
We list their masses, dominant decay modes and recall the most suitable 
reactions for the detection. We compare the results with generic one-family,
QCD-like model~\cite{FarhiSusskind,variousTC,TCsignatures,EHLQ} and walking 
technicolor~\cite{walkingTC}. This is by no means an exhaustive survey of 
PGBs in TC scenarios. We take two examples to show the similarities and stress 
the differences of TC-GIM models.

Color octet particles in TC-GIM models receive the dominant contribution to 
their masses from the one-gluon exchange. The masses of color octet particles
can range from 190~GeV to 2~TeV if the scale $f_{S-1}$ is large, where the
dependence of the mass on the scale $f_{S-1}$ is described by
Eqs.~\ref{eq:scaleuppi} and \ref{eq:octetmass}. The lower bound comes from
the fact that leptoquarks lighter than 130~GeV are excluded, which implies
that $f_{S-1}$ is larger than 65~GeV\@. The octet particles decay into two
jets. Their production cross sections in $pp$ collisions are depicted in
Fig.~\ref{fig:octet}, the cross sections are large. However, the signature
of pair-produced octets is a four-jet signal that has a large QCD background
\cite{jetbackground}. The LHC will be able to observe color octet PGBs if 
they are lighter than 375~GeV \cite{multijet}, while lower energy hadron
colliders have no chance of discovering octet PGBs.

Color octet particles are also present in other TC models. Their signatures 
are identical to the ones of TC-GIM. However, the expected mass
range is not as large as in TC-GIM models because the scale of TC 
interactions that create PGBs is related to the scale of the electroweak 
symmetry breaking. In the one-family model, one expects the octet particles 
to have masses between 200~GeV and 400~GeV \cite{FarhiSusskind,TCsignatures}. 
In walking TC models, all PGBs receive large masses from ETC interactions. 
These interactions are characterized by a large scale due to walking. The 
octet particles in walking TC models are expected to have masses in the 
200-500~GeV range~\cite{walkingTC}. The production cross section for octet 
particles is dominated by the gluon-gluon fusion. Such process is governed 
entirely by couplings that are restricted by the gauge invariance 
(Eq.~\ref{eq:vectorcouplings}). Consequently, the production rate is almost
model independent, so the observation of color-octet PGBs does not distinguish
the models. For some choices of parameters, in walking TC models color octet
PGBs can be produced more copiously than in the TC-GIM models due to 
techni-$\rho$ meson decays into color octets~\cite{walkingTC}. If color-octet
particles are discovered one expects the observation of leptoquarks,
whose masses are 1.5 times smaller than that of octet PGBs.

Color triplet particles in TC-GIM models also receive dominant mass 
contribution from the one-gluon and one-photon exchanges. Their masses are 
bounded by Tevatron experiments to be larger than 130~GeV \cite{D0,CDF}, while
the upper limit is as high as 2~TeV\@. The masses of leptoquarks are described
by Eq.~\ref{eq:scaleuppi}. Leptoquarks can be pair-produced in $e^+e^-$
colliders and discovered up to the kinematic limit. However, since they are
expected to be heavy, the best place for their discovery is a hadron collider.
The production cross section in $pp$ collisions is presented in
Fig.~\ref{fig:lq}. The upgraded Tevatron will be able to discover leptoquarks
up to 440~GeV and the LHC up to 1160~GeV\@.

As we have already described, the right-handed parts of the down quarks and the
charged leptons transform under different ETC group than the right-handed
parts of up quarks and neutrinos. Similarly, there are separate copies
of the light fermions, one copy for the up sector and one for the down sector.
This separations of the sectors is visible from the moose diagram in
Fig.~\ref{fig:model}. Consequently, there are two types of PGBs. The up-type
bosons decay into charge 2/3 quarks and/or neutrinos, the down-type into
charge 1/3 quarks and/or charged leptons. This feature is important in case
of leptoquarks. TC-GIM models have two types of leptoquarks, one whose decay
products always contain a charged lepton and the other type with a neutrino
among its decay products. All our remarks about leptoquarks that we made so
far apply to the down-type leptoquarks. The up-type leptoquarks are more
difficult to observe, since their signatures are a hadronic jet and missing
energy. Leptoquarks in other technicolor models can have various branching
ratio for decays into neutrinos and charged 
leptons~\cite{FarhiSusskind,TCsignatures}. 

The one-family model predicts leptoquarks in the range of 150--350~GeV, while
walking TC models between 200 and 500~GeV\@. The discovery limit depends on
the details of the model---the branching ratio of the decays into a charged 
lepton and a jet---and that limit is generally smaller than in TC-GIM models.
Like in the octet case, the rates for pair production in hadron collisions 
are almost model independent. If leptoquarks are discovered, the measurements
of the branching ratio into charged leptons and the masses of leptoquarks can
give some hints about viable models. Moreover, in most TC models, the 
couplings of PGBs to the ordinary fermions have much larger magnitudes such 
that the leptoquarks can be singly-produced with an associated fermion pair.
The expected rates of the production of single PGBs in TC-GIM models are
negligibly small. 

Color neutral particles are very light in TC-GIM models. Their masses range 
from 0.1 to 5 GeV as described in Eq.~\ref{eq:highmass} and Table 1. In the 
lowest order chiral perturbation theory, their masses do not depend on the
scale $f_{S-1}$. In QCD-like models one expects the masses of color-singlet 
PGBs to be in the range of 4 to 40~GeV \cite{TCsignatures}. The ETC 
interactions in walking TC models have large contributions to the masses of 
PGBs, which are between 100 and 350~GeV \cite{walkingTC}. The majority of 
neutral PGBs in the TC-GIM models are unobservable in any existing or planned
experiment because of the very weak couplings to ordinary fermions. The only
neutral particle with anomalous couplings to gauge boson pairs is the $P^0$.
Its mass is about 1~GeV and it decays most likely  into a small number of 
pions or a $\mu^+\mu^-$ pair if $f_{S-1}$ is larger than 200~GeV\@. The $P^0$
can be produced in hadron collisions, but it does not stand out from the 
hadronic background. The best environment for the discovery of $P^0$ are 
$e^+e^-$ colliders using the $e^+e^-\rightarrow P^0 \gamma$ reaction. The
production cross sections for this process are given by Eqs.~\ref{eq:ztopgamma}
and \ref{eq:eetopgamma}. Since the $P^0$ mass is so small and lies in a narrow
range around 1~GeV, the discovery potential depends not on the mass but on the
strength of the anomalous coupling. The magnitude of the anomalous coupling is
inversely proportional to $f_{S-1}$ (Eq.~\ref{eq:anomgeneral}), thus the 
production rates are proportional to $\frac{1}{f_{S-1}^2}$. The higher the 
collider's energy and luminosity, the larger are the values of $f_{S-1}$ that
can be probed. The LEP collaborations have collected enough data to probe 
$f_{S-1}$ up to 80~GeV (not all the data has been analyzed), while the NLC 
will be able to probe the scale of $SU(S-1)$ interactions up to 390~GeV\@. 

In QCD-like and walking TC models, there is usually a larger number of neutral
PGBs that exhibit anomalous couplings~\cite{variousTC,TCsignatures,walkingTC}.
These PGBs can be searched for in the same channels as the TC-GIM models: 
$e^+e^-\rightarrow P \gamma$ and $e^+e^-\rightarrow P Z^0$. QCD-like models 
generally predict low production cross sections \cite{PGBatZpole}.
The production rates depend on the number of technicolors, the technicolor
scale and the anomaly coefficient. The rates are very close to limits LEP
can place. For some models, PGBs might escape detection at LEP due to
small rates. Only the NLC will provide sufficient energy and
luminosity to observe neutral PGBs in the whole range of expected masses.
Compared to the one-family model, the walking TC models are characterized
by several low scales which greatly enhance the production rates. LEP2 has a
chance of observing walking-TC bosons if their masses are about 100~GeV\@.
The NLC with $\sqrt{s}=500$~GeV can discover neutral PGBs as heavy as 
350~GeV~\cite{walkingTC}. Obviously, the decay modes of PGBs in QCD-like
and walking TC models depend on their masses. Bosons lighter than 100-150~GeV
decay predominantly into $b\bar{b}$ pairs, while heavier particles into 
$\gamma\gamma$ pairs.

Another feature that might help distinguish the different TC scenarios 
are PGBs that are color neutral but carry the electric charge. Such PGBs
do not exist in TC-GIM models, although many other TC models predict them.
Due to the coupling to the photon, they can be pair-produced in $e^+e^-$ 
colliders and discovered almost up to the kinematic limit. Future experiments 
may easily exclude models which predict such particles with masses that are
too low.

A strong support for the TC-GIM models would be the observation of several
types of PGBs predicted by the model. One could then test the ratio of the 
octet PGBs masses to the masses of the leptoquarks. This ratio should be very 
close to 1.5, which reflects the fact that the dominant contribution to
the masses comes from the one-gluon exchange diagrams. The ratio of octet
to triplet masses provides an indirect estimate of the size of ETC 
contributions to the masses of PGBs. Smaller ratio indicates large ETC 
contributions. For instance, the walking TC scenarios \cite{walkingTC} predict
this ratio to be around one. A ratio smaller than 1.4 would rule out TC-GIM
models in their present form. Once the masses of the leptoquarks are
measured, next goal would be the measurement of the $P^0$ production rate
in $e^+e^-$ collisions.  Both the masses and the production rates depend
on $f_{S-1}$, so it would be possible to check if they yield consistent 
values of $f_{S-1}$.
    
\section{The $f_{S-1}$ scale}
\label{sec:scales}
In this section we discuss the scale of $SU(S-1)$ interactions. We describe
theoretical constraints on that scale, summarize current experimental 
limits and the discovery reach of future colliders. We also comment on the 
possibility that the exotic fermions are heavier than $f_{S-1}$. In such 
a case fermions do not condense. Instead of forming PGBs they form mesons 
resembling heavy-quark systems.

The scale of $SU(S-1)$ interactions is to a large extent a free parameter
of the TC-GIM models. As long as this scale is somewhat below the scale
of ETC interactions, it does not affect the pattern of symmetry breaking.
Thus, all one can expect is that $f_{S-1}$ is smaller than about 1000~GeV
\cite{themodel,physfromvacuum}. In some models, $f_{S-1}$ can be limited 
to a much smaller value. An example being the low-scale model, where
condensates of fermions transforming under $SU(S-1)$ group contribute
to the masses of ordinary fermions. The ETC interactions create four-fermion
operators of the form
\begin{displaymath}
  \frac{v_{TC}^2}{f_{ETC}^4} \, (\bar{q_L} q_R)\, (\bar{Q_L} Q_R),
\end{displaymath}
which involve ordinary fermions $q$ and exotic fermions $Q$\@.
When $Q_L$ and $Q_R$ form condensates, such operators contribute to 
the masses of quarks and leptons by $\frac{4 \pi f^3_{S-1} v_{TC}^2}
{f_{ETC}^4}$. In the low scale model, mass contribution from the $SU(S-1)$
condensates is identical for all down-type and all up-type fermions,
thus it should not exceed electron mass. This means that $f_{S-1}$  
cannot be larger than 15~GeV~\cite{themodel}.

The high-scale models avoid this limitation by arranging the fermion
condensates such that they do not contribute to ordinary fermion masses.
The high-scale model depicted in Fig.~\ref{fig:highscale} leaves the vacuum
alignment to be determined by the strong dynamics. It is not impossible 
that the way condensates form depends on fermion flavor. Some flavors may 
form condensates of the form $[n+12_L,S-1]$ with $[S-1,A]$ and  
$[A,S-1]$ with $[S-1,n+12_D]$, these do not contribute to the masses of
ordinary fermions. Other flavors may form condensates $[n+12_L,S-1]$ with 
$[S-1,n+12_D]$ and $[A,S-1]$ with $[S-1,A]$. It is not a disaster if such
condensates form for the top quark; such condensates do not limit $f_{S-1}$
because the top quark is so heavy. Large mass contributions originating 
from $SU(S-1)$ condensates could explain the fermion mass hierarchy 
and at the same time make it possible to have a larger scale of ETC 
interactions. Such contributions are proportional to the cube of $f_{S-1}$
and are of the order of the top quark mass only if $f_{S-1} \sim 1\ {\rm TeV}$.
Numerically, the contribution equals 
$150\ {\rm GeV} (\frac{f_{S-1}}{1\ {\rm TeV}})^3$.

\begin{figure}[t]
  \PSbox{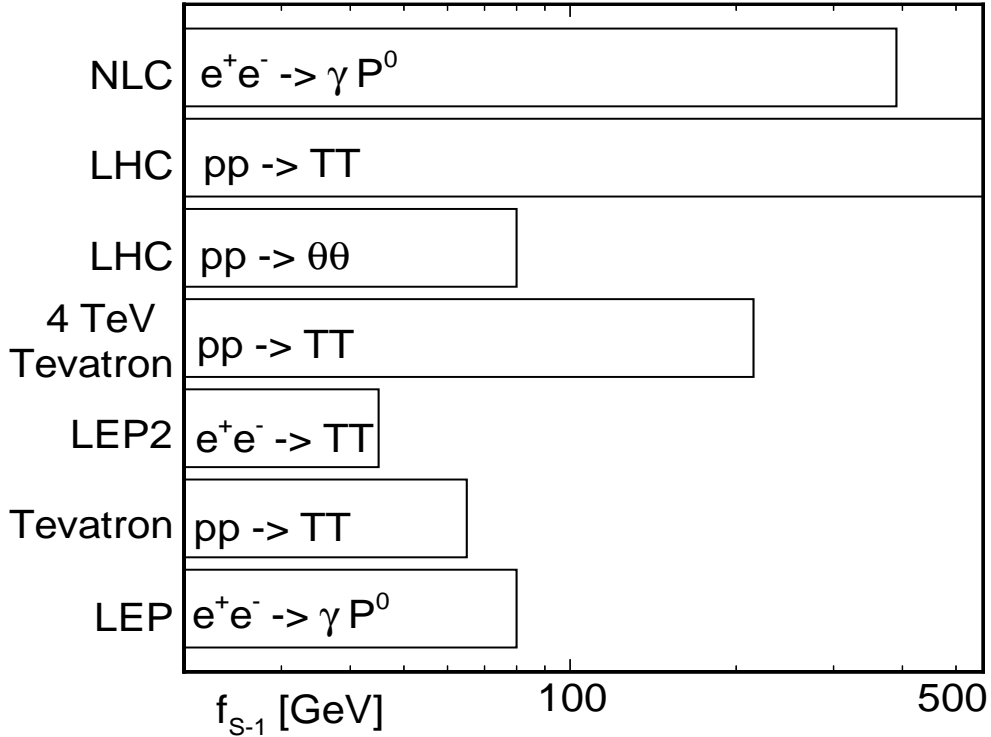 hscale=80 vscale=75 hoffset=-10 
         voffset=-215}{13.7cm}{10cm} 
  \caption{The potential of probing the scale of $SU(S-1)$ interactions at
           present and future colliders (the high-scale models). The most
           important reaction(s) for each collider is(are) indicated inside
           the bars.}
  \label{fig:scales}
\end{figure}

Lack of experimental evidence for new fermions or light PGBs imposes lower
limits on $f_{S-1}$. We summarize the reach of various experiments 
in Fig.~\ref{fig:scales}. Presented results apply to the high-scale
models of Figs.~\ref{fig:highscale} and \ref{fig:highscaleab}a. Currently,
the best limits come from the Tevatron experiments, where one places a lower
bound of 130 GeV on leptoquark masses. Thus, $f_{S-1}$ must be larger than 
65~GeV\@. The LEP experiments have not observed $P^0$, which places a limit
of 38~GeV on $f_{S-1}$. However, this result has not been updated yet with
all the data collected up to know. If all the data were analyzed, LEP could
probe $f_{S_1}$ up to 80~GeV, which would be the most competitive result
available at present. The future proton colliders can greatly enhance 
leptoquark mass limits. The $\sqrt{s}=4$~TeV Tevatron will probe $f_{S-1}$
up to 215~GeV, and the LHC up to 560~GeV\@. LEP2 is not likely to provide
any interesting information about $SU(S-1)$ interactions. One cannot fully 
take advantage of having energy larger than the $Z^0$ mass due to the small 
cross sections outside the $Z^0$ resonance peak. Limits comparable to the LHC
discovery reach can be obtained by a high-energy $e^+e^-$ collider. The NLC
will be able to probe $f_{S-1}$ up to 390~GeV by searching for the process 
$e^+e^- \rightarrow \gamma P^0$.

The current limits and discovery potential are not very different for the 
high-scale model of Fig.~\ref{fig:highscaleab}b. Usually the limits are at most
a few GeV lower than  the limits presented in Fig.~\ref{fig:scales}. However,
there is an exception. The process $p p \rightarrow P^0 X$ can be a very
sensitive probe, depending on the value of $P^0$ mass. In this model, $P^0$
can be much heavier than 1~GeV, and consequently, easier to detect. 
Irrespectively of the mass, the $P^0$ should be distinguishable from the 
SM Higgs boson. $P^0$ has large branching ratio of the decays 
$P^0\rightarrow \gamma \gamma$ of about 2\%. Unlike the Higgs boson, 
it decays mostly into light mesons, due to large $P^0\rightarrow g g$ 
decay width. 

It is also possible that at least some of the fermions are heavier than
the scale of $SU(S-1)$ interactions. This might happen in the model of 
Fig.~\ref{fig:highscale} in case of misaligned exotic top quark. Also,
in the model of Fig.~\ref{fig:highscaleab}b, such a possibility exists 
if the operators providing masses to the exotic fermions are much larger
than expected. In such a case, fermion condensates do not form, and observable
particles are no longer Goldstone bosons of spontaneously broken symmetry.
Mass terms are large enough, so that the chiral symmetry is explicitly
broken at the scale where $SU(S-1)$ interactions become confining.

This situation essentially resembles heavy quark case in QCD. Heavy 
exotic fermions decay via four-fermion interactions connecting ordinary 
and exotic fermions. The operators responsible for their decays have the 
familiar form of current-current interaction
\begin{displaymath}
  \frac{1}{f_{ETC}^2} \, (\bar{q_1}\gamma^\mu (1\pm \gamma^5) q_2)
                      \, (\bar{Q_2}\gamma_\mu (1\pm \gamma^5) Q_1),
\end{displaymath}
where $q_i$ are ordinary fermions and $Q_i$ the exotic ones. A heavy exotic
fermion $Q_1$ decays into a lighter exotic fermion $Q_2$ and a pair
of ordinary fermions. The lightest exotic fermions cannot decay this 
way. Depending on their masses they either form PGBs, as we have described 
in detail, or heavy meson states that are singlets under $SU(S-1)$ 
interactions. Heavy mesons decays are mediated by the same four-fermion 
operators; here two exotic quarks annihilate into two ordinary fermions.
Such heavy fermions can be searched for by means similar to searches
for the fourth generation quarks and leptons. The lack of weak $SU(2)$ 
interactions does not play any important role in hadronic experiments.
Exotic quarks are excluded up to masses comparable to the mass of the 
top quark. How do we know that the recently discovered~\cite{topquark}
top quark is not an exotic quark? The top quark decays into a real $W^\pm$,
which would not be the case of exotic quarks. LEP experiments exclude both 
exotic quarks and leptons up to half of the $Z^0$ mass.

\section{Conclusions}
We investigated the phenomenology of realistic technicolor models that 
incorporate the GIM mechanism. The PGBs of the TC-GIM models are not formed
by the technifermions, unlike in old technicolor theories or more realistic
walking technicolor scenarios. Anomaly-free models require additional
fermions at low-energy scales, below the scale of ETC interactions.
These extra fermions form PGBs, which are the lightest new states in the
TC-GIM models.

We have described and studied several possible realizations  of the 
light-fermions sectors. The spectrum of the PGBs is very rich. In all cases, 
among the PGBs there are leptoquarks, color octet particles and color and 
charge neutral states. Despite the fact that couplings to fermions are 
relatively small, all PGBs are short-lived particles, with lifetimes small
enough not to escape detection.

Hadron colliders are most suitable for studies of leptoquarks. Color
octet particles are much more difficult to observe due to too large QCD 
background, even though they are produced more copiously than leptoquarks.
By studying the processes $ p p \rightarrow \theta \theta $ and 
$ p p \rightarrow T T$, the LHC experiments can discover octet particles
lighter than 375~GeV and leptoquarks lighter than 1160~GeV\@. Currently,
the best limits are placed by experiments at Tevatron, which exclude
leptoquarks lighter than 130~GeV\@.

Electron colliders are capable of studying both leptoquarks and $P^0$
production. However, electron colliders do not have large enough energy
to contribute significantly to leptoquark searches. In TC-GIM, cross sections
for single production of leptoquarks are too small to yield observable 
rates. This limits the discovery reach of $e^+e^-$ colliders to half of
the CM energy. $P^0$ is the only neutral particle that can be produced with 
large enough rates, which result from the anomalous couplings of $P^0$. 
The rate for the process $e^+e^-\rightarrow P^0 \gamma$ depends on 
the strength of the anomalous coupling, so the rate is sensitive to 
$f_{S-1}$. At present, LEP excludes $f_{S-1}$ smaller than 38~GeV, but
after analyzing all the data collected so far it can probe this scale up to 
80~GeV\@. Significant improvement can be achieved at NLC, which can test 
$f_{S-1}$ up to 390~GeV.

A new feature of TC-GIM models is a very light $P^0$, with mass around 
1~GeV\@. The mass of $P^0$ does not depend on the scale of $SU(S-1)$ 
interactions. This is a consequence of the flavor symmetries of the 
high-scale models. Another prediction of the TC-GIM models is the fact that
masses and interactions of the PGBs depend on very few parameters---scales
of confining interactions. Various colliders can independently probe those
scales. Once a new signal is discovered it will be easy to check if that
signal supports TC-GIM models, since such a large number of particles is 
predicted. Planned colliders, from upgraded Tevatron to LHC and NLC have 
a chance of exploring large range of the allowed parameter region, and quite
big chances of finding signs of PGBs.   

\section*{Acknowledgments}
I would like to thank Lisa Randall for suggesting this investigation
and discussions.
\newpage


\end{document}